\documentclass{desyproc}

\begin{document}
\title{Status of the BMV experiment}

\author{{\slshape  Franck Bielsa$^1$, R\'emy Battesti$^1$, Mathilde Fouch\'e$^{2,3}$,  Paul Berceau$^1$, C\'ecile Robilliard$^{2,3}$, Gilles Bailly$^{2,3}$, S\'ebastien Batut$^1$, Julien Mauchain$^1$, Marc Nardone$^1$, Oliver Portugall$^1$, Carlo Rizzo$^{2,3}$}\\[1ex]
$^1$Laboratoire National des Champs Magn\'etiques Intenses
(UPR 3228, CNRS-INSA-UJF-UPS), F-31400 Toulouse Cedex, France.\\
$^2$Universit\'e de Toulouse, UPS, Laboratoire Collisions Agr\'egats
R\'eactivit\'e, IRSAMC, F-31062 Toulouse, France.\\
$^3$CNRS, UMR
5589, F-31062 Toulouse, France.}

\contribID{lindner\_axel}

\desyproc{DESY-PROC-2009-05}
\acronym{Patras 2009} 
\doi  

\maketitle

\begin{abstract}
In this contribution we present the status of the BMV experiment whose goal is to measure the vacuum magnetic birefringence.
\end{abstract}

\section{Introduction}

In this contribution we present the status of the BMV (Bir\'efringence Magn\'etique du Vide) experiment \cite{Battesti}
whose goal is to measure the vacuum magnetic birefringence i.e. the birefringence induced in vacuum by the presence
of an intense magnetic field. A linearly polarized light passing through a region where a magnetic
field $B$ perpendicular to the direction of propagation is present will acquire an ellipticity $\Psi$ because of the
vacuum magnetic birefringence. $\Psi$ can be written as $\Psi=\frac{\pi}{\lambda}\Delta n B^2 L$, where $\lambda$ is the light wavelength, $\Delta n$ is the difference between the index of refraction of the light polarized parallel
to the magnetic field and the index of refraction of the light polarized perpendicular to the magnetic field and $L$ is the length of
the magnetic field region. When $B$ is given in Tesla $\Delta n$ is expected to be about $4\times 10^{-24}$. It is clear looking to previous equation that the critical parameter for experiments looking for vacuum magnetic birefringence is $B^2 L$.
Our choice, since the beginning in 2001, has been to reach a $B^2 L$ approaching 1000 T$^2$m having a $B$ as high as possible with a $L$ as small as possible
to set-up a table-top optical experiment which, we believe, has the best chances of success.

A value of $\Delta n$ has been first calculated in the seventies \cite{BB} starting from the Heisenberg-Euler Lagrangian established in 1935
\cite{Heisenberg-Euler} to describe the photon-photon interactions in the framework of the Quantum ElectroDynamics based on the Dirac's model of vacuum. At the lower orders in $\alpha$, the fine structure constant, $\Delta n$ can be written as

\begin{equation}\label{Deltan}
\Delta n =\frac{2}{15}\frac{\alpha^2\hbar^3}{m_{e}^4c^5}(1+\frac{25}{4\pi}\alpha) \frac{B^2}{\mu_0}
\end{equation}

where $\hbar$ is the Planck constant over 2$\pi$, $m_e$ is the electron mass, $c$ the speed of light in vacuum,
and $\mu_0$ is the magnetic constant. The $\alpha^2$ term is given in ref. \cite{BB}. The $\alpha^3$ term has been first reported in ref. \cite{Ritus} and
it corresponds to the lower order radiative correction to the main term.
Its value is about $1.5 \%$ of the first order term.

Using the 2006 CODATA recommended values \cite{FuCo}
for the fundamental constants, equation (\ref{Deltan}) gives $\Delta n = (4.031699 \pm 0.000005)\times10^{-24} B$(T)$^2$.


As we see, the error due to the knowledge of fundamental constants is negligeable compared with the error coming from the fact that only first
order QED radiative correction has been calculated. QED $\alpha^4$ radiative correction should affect the fourth digit and the QED $\alpha^5$ radiative correction
the sixth digit. Thus measurement of $\Delta n$ up to a precision of a few ppm remains a pure QED test.

It is known that an experiment designed to measure the vacuum magnetic birefringence can also give limits on the mass and
the coupling constant of pseudoscalar particles that couples with two photons like axions or axionlike particles \cite{BRFT}. Using formulas
given for example in ref. \cite{BRFT} on the ellipticity induced because of the existence of axionlike particles  on polarized light propagating
in  the presence of a transverse magnetic field, one can infer the limits that can be given by an experiment like BMV \cite{Battesti}.
If we suppose that the QED test has been succesfully performed e.g. the $\Delta n$ value given previously has been found experimentally
with a precision of $10\%$, a coupling constant lower than about $5\times 10^7$ GeV will be excluded for axionlike particles of a mass between $10^{-3}$ and $10^{-2}$ eV.
A QED test of a precision of $1\%$ i.e. at the level of the first radiative correction, will give a coupling constant limit of a few $10^8$ GeV. Vacuum magnetic birefringence experiments are intrinsically different from astrophysics searches because they are terrestrial and all the experimental parameters are under control. Their discovery potential is therefore important. On the other hand, as far as we understand, their capacity to give interesting limits is reduced, and unless quoting unreasonable precisions in the QED test i.e. the measurement of $\Delta n$, astrophysical existing limits \cite{CAST} are already better than the ones that can be obtained via the vacuum magnetic birefringence.

The BMV experiment is a collaboration between the Laboratoire Collision Agr\'egats R\'eactivit\'e (LCAR) of Toulouse, France \cite{LCAR}, the Laboratoire National Champs Magn\'etiques Intenses (LNCMI) of Toulouse and Grenoble, France \cite{LNCMI}, and the Laboratoire Mat\'eriaux Avanc\'es (LMA) of Lyon \cite{LMA}, France. The apparatus is set up at the Toulouse site of the LNCMI, which is a laboratory specialized in pulsed magnetic fields.

To produce very high magnetic fields the only way is to have a strong current flowing into a coil. There are two main problems : heating and magnetic pressure. Pulsed fields have the advantage compared with continous fields that coils do not have the time to heat and very high fields can be reached. On the other hand, magnetic pressure which is proportional to $\frac{B^2}{\mu_0}$ becomes very important and ultra strong conductors and special external reinforcement are needed.
At the LNCMI of Toulouse, thanks to a 14 MJ bank of capacitors, pulsed fields of more than 80 T has been obtained using coils of solenoid type.

\section{Present status}

The BMV experiment is detailed in ref. \cite{Battesti}. We need a transverse magnetic field, since 2002 we have designed and tested a new geometry of coils, namely X-coil, which have reached more than 14 T over 0.25 m length corresponding to a 28 T$^2$m. Coils are operated at liquid nitrogen temperature and repetition rate is about 5 pulses per hour.
In 2006 a clean room to host the experiment was realised. To increase the optical path of light in the magnetic field region we developed a Fabry-Perot cavity to which a laser is locked.
We plan to use for such a cavity mirrors provided by LMA. Mirrors losses are of the order of a few ppm and they always have to be kept in a dust free environment. A first version of the experiment mounted in our clean room consists of two Xcoils with their cryostats surrounding the vacuum pipe and four vacuum chambers where polarizing prisms and cavity mirrors are located. Optics under vacuum is all placed on a 3.6 m table which satisfies our project requirement to have a table top experiment.
As for data analysis, at the exit of the cavity we measure both the light intensity corresponding to light polarized like the light entering in the magnetic field region $I_t$ and the light intensity corresponding to light polarized perpendicularly to the polarization of the light entering in the cavity $I_{ext}$. The ellipticity to be measured $\Psi(t)$ can be written as

\begin{equation}\label{Psi}
\Psi(t)=\Gamma \sqrt{1+\frac{I_{ext}-I_t(\sigma^2+\Gamma^2)}{I_t\Gamma^2}}-\Gamma
\end{equation}

where $\Gamma$ is the ellipticity due to the cavity, and $\sigma^2$ is the polarizer exctinction. When no magnetic field is present, and therefore $\psi(t) = 0$ one can obtain the value of $\Gamma$ as a function of $\sigma^2$, $I_t$ and $I_{ext}$. Since in principle $\Psi(t)=kB^2(t)$, for each pulse we calculate the correlation between $\Psi(t)$ and $B^2(t)$, and finally a statistical analysis will give the mean value of $k$ and its error.

We have recently taken data to measure Cotton-Mouton effect of different gases \cite{RizzoRizzo}, like air, molecular nitrogen and helium. The Fabry-Perot cavity used for such measurements is 2.2 meter long, corresponding to a free spectral range of 68 MHz. Once the laser is locked to the cavity we infer the cavity finesse by the measurement of the intensity decay time $\tau$ following a sudden stop of the light entering in the cavity. Typically $\tau$ is about 300 $\mu$s corresponding to a finesse of about 130 000, a cavity linewidth of about 520 Hz (FWHM), and a quality factor $Q$ of about 5.4 10$^{11}$. Magnetic pulse duration is about 4 ms, which is comparable with the photon lifetime in the cavity.
Actually, we have observed that ellipticity pulse is deformed by the cavity acting as a low pass filter of about 260 Hz cut-off frequency, as predicted in ref. \cite{Battesti}.
In figure 1 we show the raw data corresponding to the Cotton-Mouton effect of Helium gas, which is the smallest that one can find in nature (except vacuum effect).

\begin{figure}
\centerline{\includegraphics[width=0.5\textwidth]{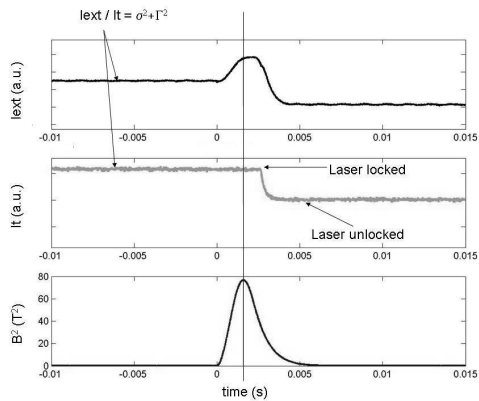}}
\caption{1 atm Helium Cotton-Mouton effect : raw experimental
data}\label{HeliumBMV}
\end{figure}

We have performed some measurements at different pressure between 0.1 to 1 atm. We have obtained a preliminary value for the $\Delta n$ per Tesla of a 1 atm of Helium gas of $\Delta n = (2.1\pm0.4)\times 10^{-16}$ in agreement with the theoretical prediction $\Delta n = 2.4 \times 10^{-16}$ and the other three experimental values published \cite{Helium}. We have also performed measurements in vacuum that are compatible with zero within the errors. Thanks to the pulse duration our frequency working point is around 500 Hz, current sensitivity is about 10$^{-7}$ 1/$\sqrt{Hz}$, mostly limited by the photodiode noise equivalent power. We are upgrading the detection system. We are also working to decrease $\Gamma$ and $\sigma^2$ which also limit the sensitivity.

\section{Short term and long term pespectives}
In the near future, we plan to measure helium Cotton-Mouton effect precisely with a sensitivity better than $\Delta n = 10^{-19}$ per Tesla per pulse, which also will give a precise calibration of our instrument. Vacuum measurements will follow. Using LMA mirrors (the expected cavity finesse is about 600 000) we plan to reach $\Delta n = 10^{-22}$ per Tesla and therefore to give new terrestrial limits on the oscillations of photons into massive particles
in 2010.

Long terms perspectives depend on the possibility to have higher magnetic fields. We have designed a new pulsed coil, namely XXL-Coil, which should reach a field higher than 25 T when a current higher than 27 000 A is injected. An XXL-Coil should provide more than 200 T$^2$m. One XXL-Coil is under construction, and winding started in july 2009. Test will follow as soon as possible. Eventually, three of them will be installed in the final set up. We hope in the next few years to finally reach our goal that is to measure the vacuum magnetic birefringence.\\

This work is supported by the {\it
ANR-Programme non th\'{e}matique} (ANR-BLAN06-3-139634), and by
the {\it CNRS-Programme National Astroparticules}.




\begin{thebibliography}{99}
\bibitem{Battesti} R. Battesti et al.,
Eur. Phys. J. D \textbf{46}, 323 (2008).
\bibitem{BB} Z. Bialynicka-Birula and I. Bialynicki-Birula, Phys. Rev. D {\bf2},  2341 (1970).
\bibitem{Heisenberg-Euler} W. Heisenberg and H. Euler, Z. Phys. {\bf38}, 714 (1936).
\bibitem{Ritus} V.I.Ritus, Sov. Phys. JETP, {\bf42}  774 (1975).
\bibitem{FuCo} Constants http://physics.nist.gov/cuu/Constants/index.html
\bibitem{BRFT} R. Cameron et al.,
Phys. Rev. D {\bf47}, (1993)  3707.
\bibitem{CAST} K.Zioutas et al. (CAST Collaboration), Phys. Rev. Lett. {\bf94}, 121301 (2005).
\bibitem{LCAR} http://www.lcar.ups-tlse.fr/?lang=en
\bibitem{LNCMI} http://lncmi.cnrs.fr/
\bibitem{LMA} http://lma.in2p3.fr/Lmagb.htm
\bibitem{RizzoRizzo} C. Rizzo, A. Rizzo and D. M. Bishop, Int. Rev. Phys. Chem. {\bf16}, 81 (1997).
\bibitem{Helium} R. Cameron et al., Phys. Lett. A \textbf{157},  125 (1991); K. Muroo et al., J. Opt. Soc. Am. B {\bf20}, 2249 (2003) ; M. Bregant et al., Chem. Phys. Lett. {\bf471}, 322 (2009).
\end{thebibliography}
\end{document}